# ARPES of Bi2212 interpreted via a particle in a system of dynamic scatterers

Udomsilp Pinsook

Department of Physics, Faculty of Science, Chulalongkorn University, Bangkok, Thailand

Email: Udomsilp.P@Chula.ac.th

**Abstract**

In this work, I employ parabolic cylinder functions in the form,

$$Ae^{-C^2z^2}D_\nu(2Cz),$$

where $z = \varepsilon^G - \varepsilon$, to quantitatively describe the ARPES spectra of an over-doped Bi2212 across the temperature range 6 - 140K at the antinode k-point. These functions come from the solutions of a particle moving in a system of random scatterers. The parameters, i.e. the overall amplitude ($A$), the spectral coherence scale ($C$), and the energy shift ($\varepsilon^G$), are determined directly by fitting to experimental data. At 140K, the dominated feature of the ARPES spectrum resembles the solution of a particle moving in a one-dimensional random system ($\nu = -\frac{1}{2}$). However, the spectrum contains some distortion, which can be fitted very well by including the contribution from $D_\nu$, with $\nu = \frac{1}{2}, \frac{3}{2}$. Between 6 – 130K, the spectra evolve in such a way that the amplitudes ($A$) of $D_{\frac{1}{2}}$ and $D_{\frac{3}{2}}$ are gradually growing with reducing temperature. As $D_{\frac{1}{2}}$ and $D_{\frac{3}{2}}$ evolve, the ARPES spectrum shows the development of depletion of states near Fermi level and of a peak-dip feature below Fermi energy. The contributions of $D_{\frac{1}{2}}$ and $D_{\frac{3}{2}}$ could be closely associated with the origin of the so-called pseudogap. Below $T_c$ (6 - 85K), the contribution of $D_1$ is emerging. The peak-dip feature becomes even more prominent. Its amplitude is continuously growing as temperature decreases, but reaches a plateau at temperature close to 6K. Furthermore, the energy shift ($\varepsilon^G$) below $T_c$ provides slightly underestimated values of the superconducting gap, compared with the measured data. Finally, the inverse of the spectral coherence scale ($1/C$), closely resembles the spectral linewidth, scales linearly with temperature above $T_c$, suggesting a connection to "Planckian" mobility in the strange-metal regime. According to the present model, the electronic states of the over-doped Bi2212 at the antinode k-point can be viewed as a realization of the theory of a particle moving in dynamic scatterers in 1D ($\nu = -\frac{1}{2}$), with corrections from the ground state solutions ($\nu = \frac{1}{2}, 1, \frac{3}{2}$) at lower temperature.

## 1. Introduction

Unconventional superconductors are a class of superconductors where BCS theory cannot give complete description, and where neither normal or superconducting state is completely understood. The unconventional superconductors exhibit richness in physics. I only give a brief overview on some relevant features that might connect to my results. In the normal state, they often exhibit the so-called pseudogap, where the electronic states contain a depletion region of DOS near Fermi level, and Planckian mobility where the resistivity is linearly proportional to temperature even at low temperature. In the superconducting state, there is no consensus on what the fundamental constituent of pairing or the pairing mechanism is. Famous examples are lanthanum barium copper oxide, discovered since 1986 [1], yttrium barium copper oxide (YBCO), discovered since 1987 [2], and bismuth strontium calcium copper oxide (BSCCO), discovered since 1988 [3]. Since then, there have been widely studied by using advanced experimental equipment, such as angle-resolved photoelectron spectroscopy (ARPES) [4-8]. The ARPES measures the true electronic dispersions of materials, i.e. the electronic states $(\varepsilon, k)$, and their intensities with associating linewidths $\Gamma$. They are the closest thing to the so-called many-body spectral function $A(\varepsilon, k)$. Indeed, to some extent, the ARPES intensity $I(\varepsilon, k)$ can be expressed by

$$I(\varepsilon, k) \propto A(\varepsilon, k) * f(\varepsilon, T), \tag{1}$$

where $f(\varepsilon, T)$ is Fermi number distribution function at temperature $T$. The intensity also convolves with equipment energy-resolution function and some random background. In addition, some parts of ARPES spectra could exhibit typical features of electron-phonon or electron-electron interactions [9]. However, some parts of BSCCO exhibit very broaden features, referred to as the "hump" part, and a peak accompanied by depletion of nearby states, referred to as the "peak-dip" part. Together, they combine or redistribute into the celebrated peak-dip-hump (PDH) spectra [10,11]. Most, if not all, of spectral analyses have used Gaussian or Lorentzian or similar peak functions to extract or capture hidden physics behind PDH. Furthermore, many theoretical works have been proposed based on the derived parameters of these PDHs in attempt to give full description to unconventional superconductors. Many theoretical frameworks conclude that single-particle regime might collapse, and some alternative regimes, such as strongly correlated many-body systems, have to be fully considered.

In this work, I perform detail investigation of the ARPES spectra from one specific example of BSCCO, i.e. over-doped Bi2212 (p = 0.21) with $T_c \approx 77K$ [8], by using the framework of a particle propagating in a system of extremely dense random dynamic scatterers. It was used for the study of heavily doped semiconductors in the past few decades. Also, the theory was formulated in terms of an effective action and solved by Feynman variational path integral. The results can give description to the band tails of heavily doped semiconductors [12]. There are largely differences between semiconductors and superconductors. Nevertheless, the solutions of the two systems share a lot of common features as I will show in the next coming sections.

## 2. Methodology

### 2.1 Brief overview of path integral method

A system of a particle interacting with $N$ random dynamic scatterers is a large subject. I give a very briefly overview of an original work by Samathiyakanit [13] (also known as Sa-yakanit). Also, there was an excellent review by Mieghem [12]. For a particle moving in a field of dense random dynamic scatterers, the action can be described by the action

$$S = \int_0^t d\tau \frac{m\dot{r}^2}{2} + \frac{i}{2\hbar} \int_0^t d\tau \int_0^t d\sigma\, W(\boldsymbol{r}(\tau) - \boldsymbol{r}(\sigma)), \tag{2}$$

where $\tau$, $\sigma$ and $t$ are time-like integral variables. $W$ is in the form of a non-local interaction from statistically averaged dynamic scatterer configurations [14]. For example, a model of Gaussian interaction can be used as

$$W(\boldsymbol{r}(\tau) - \boldsymbol{r}(\sigma)) = \frac{\alpha}{\pi^{d/2} L^d} \exp\left[-\frac{|\boldsymbol{r}(\tau) - \boldsymbol{r}(\sigma)|^2}{L^2}\right], \tag{3}$$

where $\alpha$ is the interaction strength and has the dimension of energy², $L$ is the correlation length of the random system, and $d$ is the dimension of the system. From the action, the corresponding Green's function can be derived from

$$G = \int \mathcal{D}x \, e^{\frac{iS}{\hbar}}. \tag{4}$$

For a complicated system, the action $S$ itself cannot be solved analytically. Feynman suggested a variation method,

$$G = \int \mathcal{D}x \, e^{\frac{iS_0}{\hbar}} \cdot \frac{\int \mathcal{D}x \, e^{\frac{iS_0}{\hbar}} e^{\frac{i(S-S_0)}{\hbar}}}{\int \mathcal{D}x \, e^{\frac{iS_0}{\hbar}}} = G_0 \cdot \langle e^{\frac{i(S-S_0)}{\hbar}} \rangle_{S_0}, \tag{5}$$

where $S_0$ is a suitable trial action with the corresponding Green's function $G_0$. By introducing the Bezak's action [15] as the trial action,

$$S_0 = \int_0^t d\tau \, \frac{m\dot{r}^2}{2} - \frac{m\omega^2}{4t} \int_0^t d\tau \int_0^t d\sigma \, |\boldsymbol{r}(\tau) - \boldsymbol{r}(\sigma)|^2, \tag{6}$$

where $\omega$ is a variational parameter, the approximate solutions can be derived. I give brief derivation of the solutions in Appendix. However, for deep insight, I also recommend the readers to consult Samathiyakanit's original paper [13], and subsequent works [16-19]. Bezak invented the action in Eq. (6) for an electron gas in a random potential by imitating the solution of a Brownian particle [15]. The parameter $\omega^2$ can be viewed as the inverse of the diffusion parameter. If $\omega$ is low, the particle diffusion can be large. If $\omega$ is large, the diffusion is limited, and could lead to a trapped state. The most important solution is the Green's function, which can be written as [13,19]

$$G(t) = \left(\frac{m}{2\pi i \hbar t}\right)^{d/2} \left(\frac{\omega t}{2 \sin\frac{\omega t}{2}}\right)^d \exp\left[\frac{d}{2}\left(\frac{\omega t}{2} \cot\frac{\omega t}{2} - 1\right) + \frac{i}{\hbar}\langle S \rangle_{S_0}\right], \tag{7}$$

where

$$\frac{i}{\hbar}\langle S \rangle_{S_0} = -\frac{\alpha}{2\hbar^2} \iint d\tau d\sigma \int \frac{d\boldsymbol{q}}{(2\pi)^d} W(q) e^{-B^2 q^2/2}. \tag{8}$$

The function $W(q)$ is the Fourier pair of $W(\boldsymbol{r}(\tau) - \boldsymbol{r}(\sigma))$. The factor $B^2$ is defined by Eq. (A9). In order to relate to temperature, Wick rotation can be considered explicitly as $\frac{it}{\hbar} \to \frac{1}{k_B T}$. If $T$ is high, $t$ is small. By taking the limit $t \to 0$, the solution can be reduced to the form (see Appendix)

$$G(t) \propto (it)^{-\frac{d}{2}} e^{-\frac{it}{\hbar}\varepsilon^0 - \frac{t^2}{8C^2}}, \tag{9}$$

where $C$ is just a redefinition of the factors $\alpha$, $L^2$ and $B^2$. The path integral method might look very complicated, but the solution can be viewed simply as an ensemble average of the energy phase factor $\langle e^{-\frac{it}{\hbar}\varepsilon^0}\rangle \sim e^{-\frac{it}{\hbar}\varepsilon^0 - \frac{t^2}{8C^2}}$. In addition, $C$ can be viewed as a spectral coherence scale. If $C$ is large, the term $e^{-\frac{it}{\hbar}\varepsilon^0 - \frac{t^2}{8C^2}} \to e^{-\frac{it}{\hbar}\varepsilon^0}$, and the single particle state is retained with the energy shift $\varepsilon^0$. This state becomes strongly coherent. If $C$ is small, the term $e^{-\frac{it}{\hbar}\varepsilon^0 - \frac{t^2}{8C^2}}$ dissipates quickly, and the state becomes decoherent. In the limit $t \to 0$, the solution can be considered as a very weak perturbation of free particle states because the solution is the same as of when using $S_0 = \int_0^t d\tau \frac{m\dot{r}^2}{2}$ as the trial action. This limit is also the same limit as $\omega \to 0$, where the particle diffusion can be large. In the original theory [13], $\varepsilon^0 = 0$. However, a finite value of $\varepsilon^0$ can be deliberately added in order to help adjust the band alignment. I keep the finite value $\varepsilon^0$ as well in order to ease the fitting process. Also, it would account for other residue interactions, other than the random scatterers. The density of states $\rho(\varepsilon)$ can be derived from Fourier transform of $G(t)$ as

$$\rho(\varepsilon) \propto \int dt\, (it)^{-\frac{d}{2}} e^{\frac{it}{\hbar}(\varepsilon - \varepsilon^0) - \frac{t^2}{8C^2}} = A e^{-C^2 z^2} D_{-\frac{d}{2}}(2Cz), \tag{10}$$

where parameters are the amplitude ($A$), and the spectral coherence scale ($C$), and $z = \varepsilon^0 - \varepsilon$ which accounts for the energy shift. The $D_{-\frac{d}{2}}(z)$ is the parabolic cylinder function. With the dimension parameter ($d = 1,2,3$), $\rho(\varepsilon)$ can be displayed as shown in Figure 1.

In this work, I analyze the ARPES spectra which comes from the emission of photo electrons. Thus, they provide the information from the contribution of the occupied states. Also, the energy band appears to simply runs down from Fermi surface. Therefore, the energy variable is in a reverse sign, i.e. $E = -\varepsilon$. After applying the reverse sign, some special features of these functions can be discussed. Firstly, $\varepsilon^0$ does not coincide with a peak position. As $E > 0$, all these functions resemble an exponential decay tail, i.e. $\rho(E) \propto z^\nu e^{-2C^2 z^2}$. On the other hand, for $E < 0$, $\rho(E) \propto 1/\sqrt{\varepsilon}$ ($d = 1$), $\rho(E) \propto constant$ ($d = 2$), and $\rho(E) \propto \sqrt{\varepsilon}$ ($d = 3$). These features are shown in Figure 1.

It is worth noting that there is a special case, where $d = 0$,

$$\rho(E) \propto \int dt\, e^{\frac{it}{\hbar}(\varepsilon - \varepsilon^0) - \frac{t^2}{8C^2}} = 2AC e^{-2C^2 z^2}, \tag{11}$$

which is simply a Gaussian peak, centered at $\varepsilon = \varepsilon^0$. Consequently, the inverse of $C$ closely relates to nothing but the width of the Gaussian peak, i.e. the full width at half max $FWHM = \frac{\sqrt{2\ln 2}}{C}$, which can also be related to the linewidth of the corresponding state. The Gaussian peak is introduced here for future comparison to other type of $\rho(E)$.

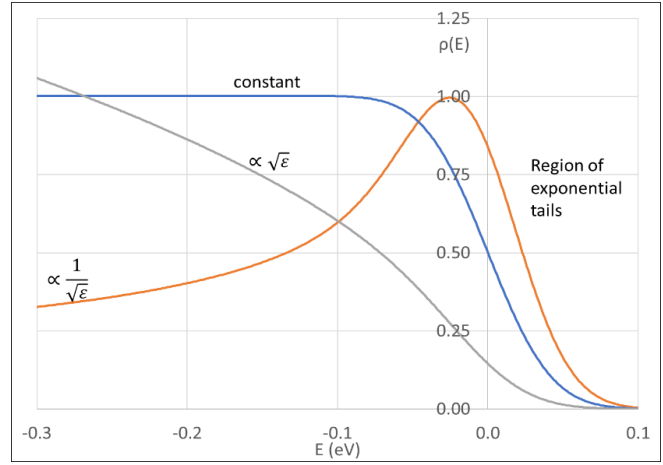

**Figure 1** shows $\rho(E)$ from Eq. (10) with $\varepsilon^0 = 0$ eV, and $C = 15$ eV$^{-1}$. The amplitudes $A$ are 0.69, 0.4 and 0.125 for $d = 1$ (orange line), 2 (blue line), and 3 (gray line) respectively. Note that $E = -\varepsilon$. Also, $\varepsilon^0$ does not coincide with any peak position.

Next, if $T$ is low, $t$ is large. Also, this limit is similar to take the limit of a large $\omega t$, as long as $\omega$ is finite. The solution is corresponding to a very strong interaction. By taking the limit explicitly, the solution becomes much more complicated [13], see Appendix, but it can be simplified into a simple form as

$$\lim_{t \to \infty} G(t) \propto (it)^{\frac{d}{2}} e^{-\frac{it}{\hbar}\varepsilon^G - \frac{t^2}{8C^2}}. \tag{12}$$

Consequently, the density of states can be obtained from Fourier transform, which can be written as

$$\rho(E) = A e^{-C^2 z^2} D_{\frac{d}{2}}(2Cz), \tag{13}$$

where $z = \varepsilon^G - \varepsilon$. I emphasize that $\varepsilon^G$ does not coincide with any peak position. This $\rho(E)$ is no longer close to those of the free particle. It composes of redistribution of states with negative and positive regions confined in an envelope controlled by $C$, as shown in Figure 2. This formula was suggested by Samathiyakanit [13] and some following works [16-19]. Samathiyakanit suggested that the expression for the effective mass cannot be identified within this limit of his theory. This would indicate that a nearly free particle picture might be failed to describe these states. Indeed, this is also the same limit as $\omega t \to \infty$, where the particle diffusion is limited, or the particle becomes localized or trapped [15]. Then, the nearly free particle's effective mass has no meaning. Moreover, there exist some regions where $\rho(E)$ composes of negative and positive values, see Figure 2. Due to their negative density of states, these expressions for $\rho(E)$ had been neglected in the past works. However, I believe that if these states are living on another system, the negative $\rho(E)$ can be interpreted as the depletion of the density of states, which is compensated by the increasing of the density of states somewhere else. The total $\rho(E)$ still has physical meaning as long as it is all positive. Also, the total number of states seem to be unchanged. These composites between positive and negative $\rho(E)$ might also be interpreted as an emergence of a trapped particle, in which the ARPES measurement can probe its existence from their net outcomes.

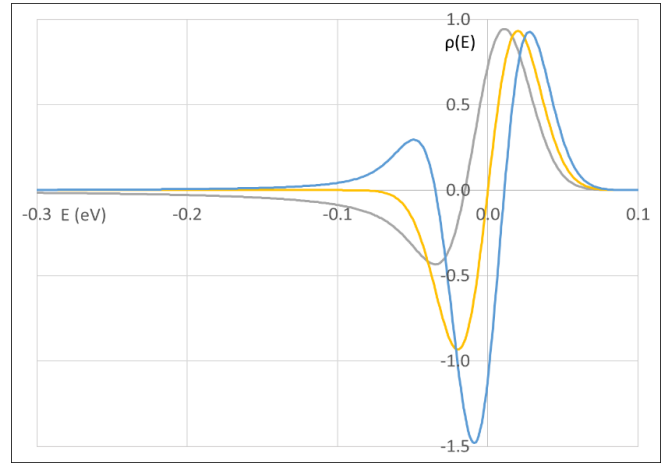

**Figure 2** shows $\rho(E)$ from Eq. (13) with $\varepsilon^G = 0$ eV, and $C = 25$ eV$^{-1}$. The scaled $A$ are 1.25, 1.54 and 1.82 for $d = 1$ (gray line), 2 (yellow line), and 3 (blue line) respectively. Note that $E = -\varepsilon$. Also, $\varepsilon^G$ does not coincide with the peak positions.

In this work, I propose that some ARPES spectra of the Bi2212 compound [8] could be viewed as a system of random scatterers. These scatterers could be oxygen vacancies, lattice distortion due to those vacancies, other random impurities, or even some fluctuating environment, such as spin fluctuation, distorted charge density wave or etc. If these scatterers can be identified, a certain theory with some predictive power can be formulated. In this work, I provide an alternative way to look at the experimental data, other than using typical peak-like functions, such as Gaussian function. It could provide some additive complementary information to some already known information. In analogy, this is similar to information from James Webb space telescope (JWST) which is complementary to Hubble's. Surprisingly, I discovered that the ARPES spectra of the over-doped Bi2212 (p = 0.21) closely resemble the shapes of the functions $\rho(E) = Ae^{-C^2z^2}D_\nu(2Cz)$, see Figure 3. For the hump part of the ARPES spectrum at 140K, it exhibits similar limiting behaviors, such as having an exponential tail at one end, and $\propto 1/\sqrt{\varepsilon}$ at another end. By subtracting the ARPES spectrum at 140K by a spectrum at lower temperature, i.e. 6K, the remaining signal looks pretty much like the functions in Figure2. Therefore, I will use this expression for $\rho(E)$ to perform fitting to the ARPES spectra of the over-doped Bi2212 between 6K-140K [8] at the antinode k-point, and let's see how far I can go.

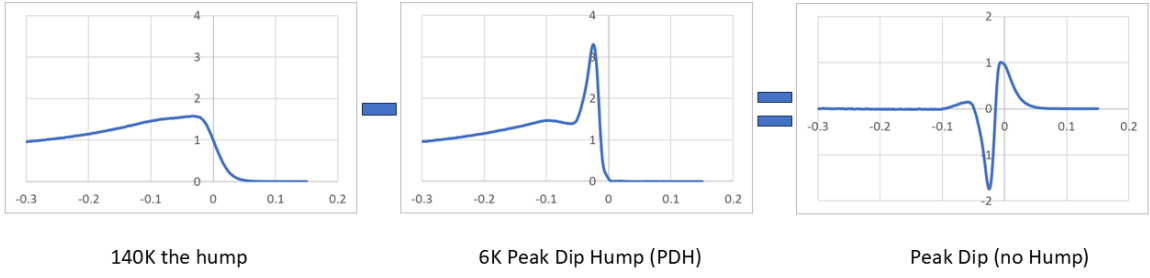

140K the hump      6K Peak Dip Hump (PDH)      Peak Dip (no Hump)

**Figure 3** shows schematic idea of how spectra can be explained by the parabolic cylinder functions. I took the data from the experiment by Chen et al. [8]. The hump at 140K is subtracted by the peak-dip-hump at different temperatures, say 6K as in the picture. The remaining signal shows the peak-dip part. Surprisingly, it closely resembles $\rho(E)$ from Eq. (13), see Figure 2 for comparison.

## 2.2 Fitting Procedure

For ARPES, the intensity can be viewed as many-body spectral function, $A(E,k)$, multiplied by Fermi number distribution, as in Eq. (1). In this work, I took the ARPES spectra from the excellent work led by Chen at el. [8]. They measured the properties of the over-doped Bi2212 (p = 0.21) with $T_c \approx 77K$ [8]. Also, the experiment set $\mu = 0$. I will drop $\mu$ from now on. The band appears to run down from the Fermi level, thus, the sign of energy variable has to be reversed, i.e. $E \to -\varepsilon$. Furthermore, I study only one k-point region, which is the so-called antinode region, i.e. $k_x \approx -0.8$ Å$^{-1}$ and $k_y$ are between $\pm 0.1$ Å$^{-1}$, according to the experimental setup [8]. For simplicity, I shall drop index $k$ from $A(E,k)$, and apply the approximation $A(E) \approx \rho(E)$. As the spectra are depended on the temperature, I will write explicitly temperature-dependent functions. Then, the ARPES intensity can be written as

$$I(\varepsilon, T) = f(E,T)\rho(E,T), \qquad (14)$$

where $f(E,T) = \frac{1}{e^{\beta E}+1}$.

    The first task is to identify any universal quantity of the system, which is invarient under the phase transition. The CuO$_2$ layer in Bi2212 compound is a quasi 2D system. Its electronic structure must exhibit some 2D features, as shown in Figure 4. It shows the ARPES intensity, Eq. (14), at 140K of the over-doped Bi2212 at 13° from the antinode k-point. Opened blue circles are the data taken from the experiments [8]. The red line is the total fitting. The blue line is the major contribution of $\rho(E,T)$ from Eq. (10) with $d=2$. The orange line is the contribution from Eq. (15), where $d=1$. The error from fitting is shown by the black line. The result suggests that the base system is dominated by the features of the 2D system with some correction from the features of the 1D system.

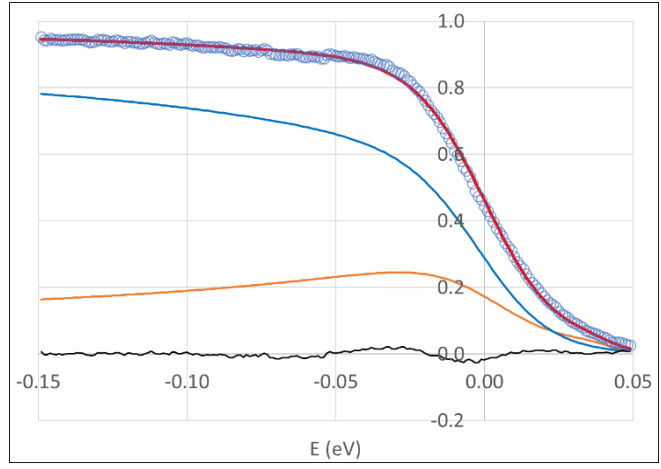

**Figure 4** shows the ARPES intensity, $I(\varepsilon, T) = f(E,T)\rho(E,T)$, at 140 K of the over-doped Bi2212 at 13° from the antinode k-point. Opened blue circles are from experiments [8]. The red line is the total fitting. The blue line is the contribution of $\rho(E,T)$ from Eq. (10) with $d = 2$. The orange line is the contribution from Eq. (15). The error from fitting is shown by the black line.

For the antinode region at 140K, I notice that the ARPES spectrum closely resembles that of a 1D system with random scatterers, where there is an exponential decay tail at one end ($E > 0$) and $\rho(E) \propto \frac{1}{\sqrt{\varepsilon}}$ at another end ($E < 0$). Thus, I propose that

$$\rho_0(E, T = 140K) = A_0 e^{-C_0^2 z_0^2} D_{-\frac{1}{2}}(2C_0 z_0), \tag{15}$$

where $z_0 = \varepsilon^0 - \varepsilon$, can be used for fitting. By the standard least square fitting, I obtain the result with fitting paramaters $A_0 = 1.266$, $C_0 = 4.639$ eV$^{-1}$, and $\varepsilon^0 = -0.093$ eV. The result is shown in Figure 5, where opened blue circles are experimental data [8], the orange line is from Eq. (15). This orange line is very important. I find that it is invariant under temperature. Therefore, it can serve as the baseline for all the fitting in this work. There are some discrepencies in the energy region around $E = 0$. The difference looks pretty much like a solution of $\lim_{t \to \infty} G(t)$, Eq. (13). These discrepencies lead us to add more terms into $\rho(E,T)$, as follows

$$\rho_1(E,T) = A_1 e^{-C_1^2 z_1^2} D_{\frac{1}{2}}(2C_1 z_1), \tag{16}$$

$$\rho_2(E,T) = -A_2 e^{-C_2^2 z_2^2} D_1(2C_2 z_2), \tag{17}$$

$$\rho_3(E,T) = -A_3 e^{-C_3^2 z_3^2} D_{\frac{3}{2}}(2C_3 z_3), \tag{18}$$

where $z_i = \varepsilon_i^G - \varepsilon$, and the total $\rho(E,T)$ can be obtained from the summation

$$\rho(E,T) = \sum_{i=0}^{3} \rho_i(E,T). \tag{19}$$

The total result $I(\varepsilon,T) = f(E,T)\rho(E,T)$ at 140K is shown as a full red line in Figure 5, 6 and 9.

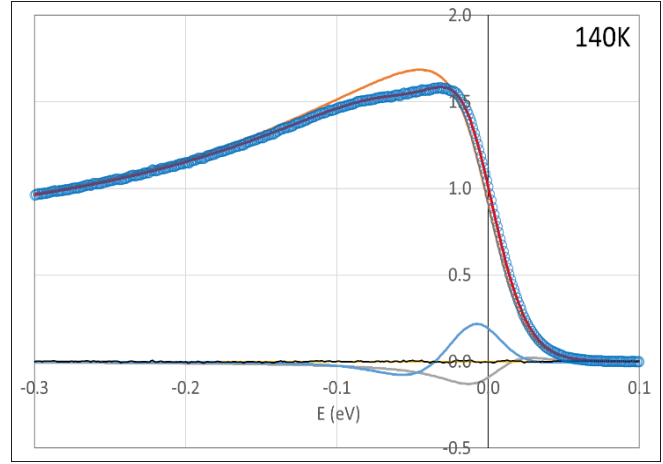

**Figure 5** shows $I(\varepsilon,T) = f(E,T)\rho(E,T)$ at 140K of the over-doped Bi2212 near the antinode k-point. Opened blue circles are from the experiments [8]. The orange line is the contribution of $\rho(E,T)$ from $\rho_0$ (Eq. (15)), the blue and gray lines are the corrections from $\rho_1$ (Eq. (16)) and $\rho_3$ (Eq. (18)), respectively, and the red line is the total fitting (Eq. (19)). The error from fitting is shown by the black line.

By using the standard least square fitting, I shall use these expressions, Eq. (14) – (19), to fit the ARPES spectra for all available temperature near the antinode k-point [8]. As $\rho_0(E,T = 140K)$ is actually invariant under temperature and also under phase transition, thus $A_0$, $C_0$, and $\varepsilon^0$ can be treated as constants. These values should reflect on the true properties of Bi2212, which should be investigated further in future works. The varying properties are those of $\rho_1(E,T)$, $\rho_2(E,T)$ and $\rho_3(E,T)$, which I shall explore further. The results of all $A_i$, $C_i$, and $\varepsilon_i^G$ as a function of temperature will be reported and analysed in the next section.

From Figure 4 and 5, they appear as if there are some dimensional crossovers in the system. The dimension-related functions seem to give a good description to the ARPES spectra. It is worth noting also that, for a single $D_\nu$, where $\nu > 0$, it contains both negative and positive densities, as shown in Figure 2. Therefore, it cannot stand alone. Rather, it could exist inside another system with large amount of density of states, so that the total density of states is always positive. The negative density could be interpreted as depletion of the total density or as an existence of a trapped state. However, the emerging negative density always accompanies by another surge in positive density, so that the total number of states seems to be conserved. The true effect on the total density of states is to redistribute the density in a limited region of energy. In addition, the function $D_\nu$ exhibits some asymmetry around its central point, which is also close to the Fermi level. If this state is real, some particle-hole asymmetry must also be detected. Indeed, this similar particle-hole asymmetry was reported before by several experiments [8,20], and I will demonstrate this effect in my data analysis as well.

## 3. Results and Discussion

### 3.1 Overview of ARPES spectra

From Figure 6, the present dataset covers the antinode spectra from 6K (Top-Left) upto 140K (Bottom-Right). Blue circles mark the raw ARPES intensity [8], while the red lines show the fitted curves, Eq. (14), together with their components $\rho_i$ from Eq. (15) - (19). The base line (orange line) from $\rho_0$ (Eq. (15)) is fixed, i.e. invariant under temperature. The overall agreement is striking, i.e. the fitting results capture the line shape at every temperature with only small, structureless residuals (black line). This gives confidence that the parabolic-cylinder framework is the correct description of these spectra.

At $T = 140K$, the baseline contribution $\rho_0$ (Eq. (15)) is already well accounted for most of the ARPES spectrum. This background remains essentially unchanged with temperature and serves as the constant reference against which all other features evolve. The changes with cooling are carried instead by the additional terms in Eq. (16) - (18), whose amplitudes sharpen and linewidths narrow in a systematic manner. It should be noted that the negative signs in Eq. (17) and (18) are not incidental. They are required for the best fits and highlight that the underlying spectral structure is more subtle than a naive additive form. In addition, the algebraic signs carry physical weight, i.e. they encode interference between components and ensure that pseudogap, coherence peak, and strange-metal dissipation can all be united within a single universal formulation.

As the behavior of $\rho_0$ (Eq. (15)) might reflect on the nature of a quasi 1D system. I would like to discuss on the nature of the Cu-O chain along the antinode k-point. At the antinode, the low-energy electronic states are dominated by the overlap of the Cu $d_{x^2-y^2}$-orbital with the O $p$-orbital along the Cu–O bond. In this geometry, the nearest-neighbor hopping parameter $t_0$ dominates the dispersion, producing the simplified tight-binding band dispersion of the form $E(k) = E_0 - 2t_0 \cos(ka)$. This cosine law is the fingerprint of a quasi-1D chain embedded in the Cu-O₂ plane. Although the lattice is two-dimensional, the orbital alignment projects a chain-like character onto a one-dimensional expression.

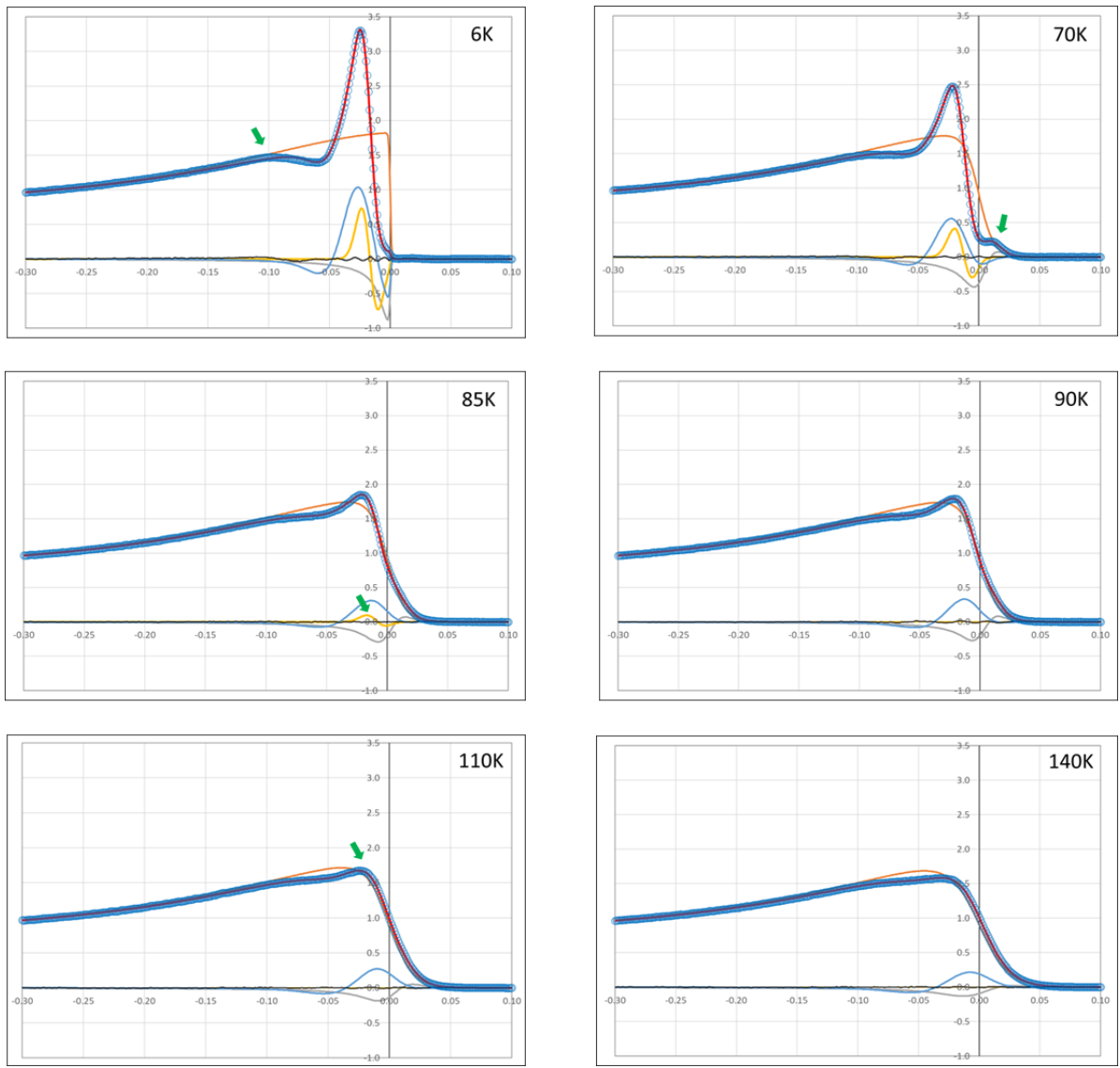

**Figure 6** shows the evolution of $I(\varepsilon, T) = f(E,T)\rho(E,T)$ as a function of temperature. Opened blue circles are from experiments [8]. The orange line shows the contribution of $\rho(E,T)$ from $\rho_0$ (Eq. (15)), blue, yellow and gray lines show the corrections from $\rho_1$ (Eq. (16)), $\rho_2$ (Eq. (17)) and $\rho_3$ (Eq. (18)), respectively. The red line is from the total fitting (Eq. (19)). The error from fitting is shown by black line. Indeed, the error at low temperature is quite noticeable. The green arrows at 6K and 70K marked some "ghost" peaks, which have no major association with any major peak in $\rho_i(E,T)$. The green arrow at 85K marked the early development of $\rho_2(E,T)$, close to superconducting phase transition. The green arrow at 110K marked the early development of the peak -dip feature. There is a short clip of this evolution at
https://drive.google.com/file/d/1263ZTSviQDIJp9C_aCpNtl_zmJDhdIvo/view?usp=sharing

Above the pseudogap phase $T > T^*$, the ARPES spectra show that the dispersive antinode band is consistent with this quasi-1D form. But as the system is cooled through the pseudogap phase $T \approx T^*$, the band dispersion changes abruptly. The antinode dispersion no longer bends with $k$. Instead, it collapses into a flat, pinned peak at nearly constant energy. This is the so-called flat band. In a tight-binding interpretation, this is equivalent to a sudden disappearance of the effective hopping, i.e. $t_{0,eff} \to 0$, not by gradual scaling but by disruption. The itinerant channel along the Cu–O bond is cut off, leaving behind localized or trapped states that no longer carry momentum information. This feature happens in a small region of $k$ around the antinode k-point, i.e. $k_x \approx -0.8$ Å$^{-1}$ and $k_y$ are between $\pm 0.1$ Å$^{-1}$, as reported by the experiment [8]. It is worth noting that the bare orbital overlap does not vanish, i.e. the Cu–O chemistry remains unchanged. What vanishes is the effective hopping seen by the spectral function. The pseudogap formation, and environmental screening suppress the kinetic channel so completely that ARPES detects no dispersion at all. In this picture, the transition to the pseudogap phase is less a gentle renormalization of parameters than a sudden projection. This would suggest that the charge transport might be limited to diffusion as the hopping channel is effectively closed. In addition, the flat band is often observed in recently discovered Kagome-type compounds [21,22]. Surprisingly, the hump signal, similar to Eq. (15), is also detected in the ARPES spectra, accompanying the flat band in these compounds [21,22].

### 3.2 Pseudogap Development

At temperatures above $T_c$ but below 140K, the ARPES spectra show a gradual depletion and redistribution of states near the Fermi level, see the area around the green arrow at 110K in Figure 6 for example. Within the present framework, this depletion is naturally accounted for by the evolution of the components $D_{\frac{1}{2}}(2C_1 z_1)$ and $D_{\frac{3}{2}}(2C_3 z_3)$ of $\rho_1$ (Eq. (16)) and $\rho_3$ (Eq. (18)), respectively.

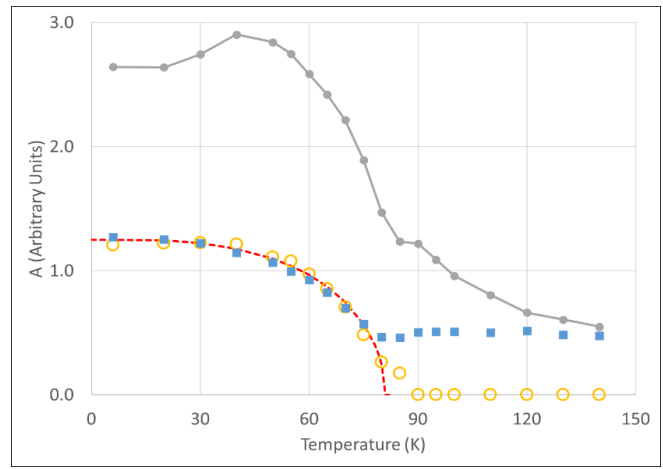

**Figure 7** shows the temperature-dependent values of fitting parameter $A_1$ (gray line and filled gray circles), $A_2$ (opened yellow circles), $A_3$ (filled blue squares). The red dashed line is calculated from the modified gap equation, Eq. (20), where I use $A(0) = 1.25$ and $T_c = 81$K, compared with the report $T_c \approx 77$K [8].

From Figure 7, the amplitude $A_1$ (gray line and filled gray circles) increases steadily while $A_3$ (filled blue squares) exists nearly constantly as the system is cooled from 140K toward $T_c$. Their evolution directly produces the spectral dip and peak near the Fermi edge — the pseudogap feature, see the green arrow at 110K in Figure 6. Unlike a true gap opening from an ordered phase, the dip part is the depletion region originated from the negative lobes of $\rho_i(E,T)$, which remove spectral weight from the vicinity of $E_F$. On the other hand, the peak part comes from the positive lobes of $\rho_i(E,T)$, which increases the spectral weight. Nevertheless, the total density is conserved. In this picture, the pseudogap is not a new order parameter but the redistribution of incoherent states as scattering becomes stronger at lower $T$.

An important signature of this interpretation is particle–hole asymmetry. As shown in Figure 2, both $D_{\frac{1}{2}}(2C_1z_1)$ and $D_{\frac{3}{2}}(2C_3z_3)$ carry some asymmetric lobes of opposite signs on different energy regions close to $E_F$, the depletion is definitely asymmetric with respect to $E_F$. In addition, at high $T$, Fermi distribution allows some states above Fermi level to be accessible. Thus, the particle-hole asymmetry around $E_F$ can be probed by experiments. There are several reports of asymmetric pseudogap spectra in the cuprate superconductors [8,20]. I demonstrate this asymmetric effect by plotting $\rho(E,T)$ (blue, red, and black lines) in both sides of $E_F$ and compare with the experiment report on $\rho(E,T) \approx \frac{I(E,T)}{f(E,T)}$ (opened circles) [8]. Without refitting, the particle-hole asymmetry comes out naturally from the model, see the behavior of the data and the model across $E_F$ in Figure 8. The discrepancies between the model and experimental data are large at larger positive $E$ because if $f(E,T)$ decays faster than $I(E,T)$, then its corresponding uncertainty will also be greatly magnified.

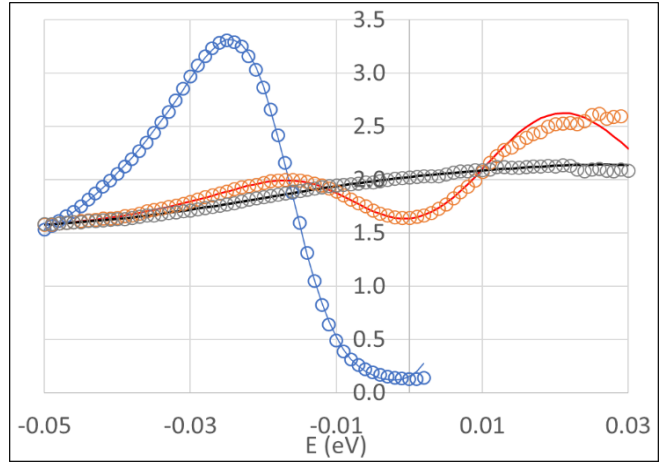

**Figure 8** show $\rho(E,T)$ and the associated particle-hole asymmetry which comes out naturally from the model. The blue, red, and black lines are from the model, Eq. (19), at 6K, 85K and 140K, respectively. Opened circles are from experimental data, where $\rho(E,T) \approx \frac{I(E,T)}{f(E,T)}$. Note that $f(E,T)$ drops faster than $I(E,T)$ at $E > 0$. Therefore, the errors could be magnified in this energy region.

### 3.3 Superconducting Coherence Peak

Below the superconducting transition temperature $T_c$, the ARPES spectra exhibit a sharp peak that grows in intensity as $T$ decreases. Within the present framework, this peak is naturally described by the contribution of emerging $\rho_2$ (Eq. (17)), associating with the existing $\rho_3$ (Eq. (18)) from the pseudogap phase. Figure 9 shows the significance of $\rho_2$ (yellow line) and $\rho_3$ (blue line) at 6K.

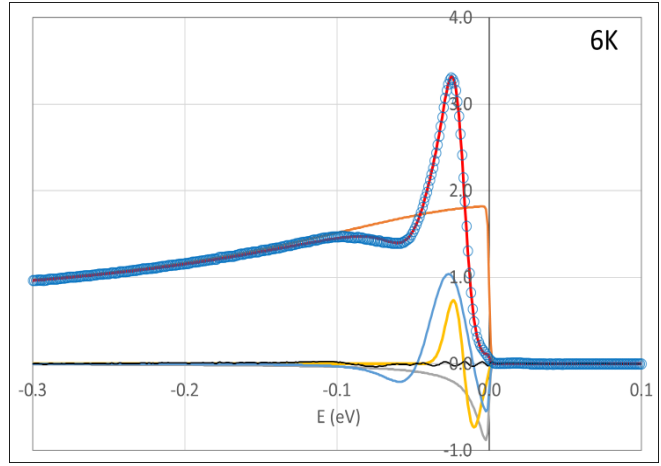

**Figure 9** shows $I(\varepsilon, T) = f(E,T)\rho(E,T)$ at 6K. Opened blue circles are from experiments [8]. The orange line shows the contribution of $\rho(E,T)$ from $\rho_0$ (Eq. (15)), blue, yellow and gray lines show the corrections from $\rho_1$ (Eq. (16)), $\rho_2$ (Eq. (17)) and $\rho_3$ (Eq. (18)), respectively. The red line is from the total fitting (Eq. (19)). The error from fitting is shown by black line. Indeed, the error at 6K is quite noticeable.

From Figure 6, the contribution of $\rho_2$ starts to grow at 85K (see green arrow at 85K), just above the superconducting phase transition. As temperature decreases, it develops in a very interesting fashion. From Figure 7, its amplitude $A_2(T)$ (opened yellow circles) shows a clear order-parameter-like evolution, i.e. it vanishes near $T_c$ and increases rapidly upon cooling, saturating at low temperature. This behavior can be well captured by the modified BCS relation (dashed red line)

$$A_2(T) = A_2(0) \tanh\left\{1.74\sqrt{\frac{T_c}{T} - 1}\right\}, \qquad (20)$$

where $A_2(0) = 1.25$ and $T_c = 81$K, compared with the reported $T_c \approx 77$K [8]. This behavior strongly suggests that the $D_1(2C_2z_2)$ term tracks the superconducting condensate.

From Figure 10, the associated energy parameters $\varepsilon_2^G$ (yellow symbols and line) and $\varepsilon_3^G$ (blue symbols and line) follow the experimental superconducting gap (red and green symbols and lines) extracted from the ARPES peak positions [8], though with a slight systematic underestimation. This is because $\varepsilon_i^G$ are not exactly the peak positions, while the experiment reported exactly the peak positions [8]. This parallel evolution indicates that $\rho_2$ (Eq. (17)), despite its origin in a disorder framework, effectively captures the emergence of coherent spectral weight tied to superconductivity.

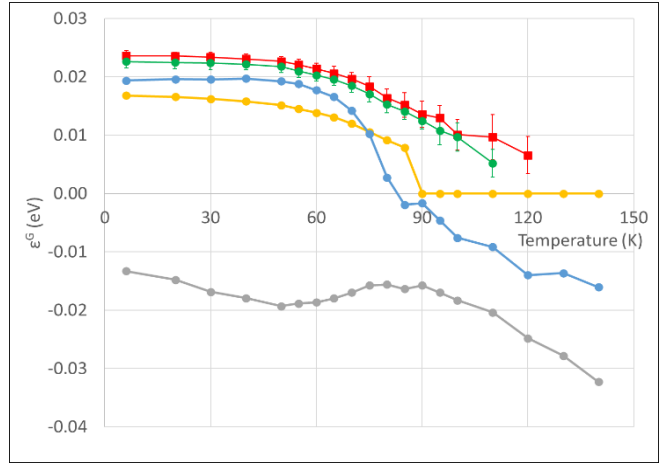

**Figure 10** shows the temperature-dependent values of fitting parameter $\varepsilon_1^G$ (gray line and filled gray circles), $\varepsilon_2^G$ (yellow line and filled yellow circles), $\varepsilon_3^G$ (blue line and filled blue circles). The experimental gap values with error bars [8] are also shown (green and red, both lines and symbols). Note that $\varepsilon^G$ are not exactly the peak positions, unlike the experimental values which reported the exact peak positions. Nevertheless, the green, red, blue and yellow lines seem to be parallel to each other to some noticeable extension.

Physically, $\rho_2$ (Eq. (17)) and $\rho_3$ (Eq. (18)) represent the redistribution of spectral weight into a sharp, coherent feature near the Fermi edge, balanced by nearby depletion. In this picture, the superconducting coherence peak is not an independent quasiparticle pole, but the natural manifestation of universal disorder states re-organization and redistribution as the system enters the superconducting phase. The remarkable agreement between $A_2(T)$ (opened yellow circles in Figure 7) and $\varepsilon_2^G$ (yellow symbols and line in Figure 10) and experimental order-parameter trends suggests that the same minimal random-scatterer framework can account for both the pseudogap above $T_c$ and the coherence peak below it.

### 3.4 Ghost peak as interference

In ARPES, a narrow peak sometimes appears in the antinode spectra and has often been interpreted as a signature of a coherent quasiparticle or a bosonic mode. Within the present framework, however, such a feature can be reproduced without invoking a new pole of the Green's function. It emerges naturally from the algebraic superposition of the signed contributions $\rho_1$ (Eq. (16)), $\rho_2$ (Eq. (17)), and $\rho_3$ (Eq. (18)) on top of the positive baseline $\rho_0$ (Eq. (15)). Each of the parabolic cylinder function $D_\nu$, where $\nu > 0$, carries some lobes with positive and negative quantities. Consequently, the total density of states, Eq. (19), exhibits redistribution that significantly differs from $\rho_0$. For $\nu = \frac{1}{2}, \frac{3}{2}$, their lobes appear to be asymmetric around their centers, as shown in Figure 2. When combined and subsequently convolved with the Fermi cutoff and instrumental resolution, these lobes may align to create a local maximum that resembles a peak. Therefore, some developing peaks might appear and then disappear as temperature changes. The example in Figure 6 is the small peak region around the green arrow at 70K which totally disappears in the next snapshot at 85K. Furthermore, some peaks are not corresponding to the peaks of $D_v$ but rather an interference of them. I shall refer to these peaks as a "ghost" peak, i.e. a striking but composite feature that arises from interference among universal $D_v$ components, but not a true pole of Green's function. The green arrows at 6K and 70K marked some ghost peaks in Figure 6. The essential point is the fragility of the ghost peak to some changes. A few-percent change in $\varepsilon_i^G$ or $C_i$ shifts the peak or eliminates it altogether, while the underlying hump $\rho_0$ remains robust.

### 3.5 spectral coherence scale and Planckian scaling

From Figure 11, it shows the spectral coherence scale $C_i$ extracted from the fitting results. Its inverse, $C_i^{-1}$, resembles the role of a scattering rate and the spectral linewidth. It is tempting to compare with the Gaussian function in Eq. (11), where FWHM $\approx 1.177 C^{-1}$. Below $T_c$, $C_i$ are quite large, and hence the term $e^{-\frac{it}{\hbar}\varepsilon_i^G - \frac{t^2}{8C_i^2}} \to e^{-\frac{it}{\hbar}\varepsilon_i^G}$. In the other word, the state becomes more coherent. As temperature increases above $T_c$, both $C_1^{-1}$ and $C_3^{-1}$ increase linearly with $T$, see the inset in Figure 11. The state becomes more decoherent. This behavior is similar to the Planckian transport where $\Gamma \approx \frac{k_B T}{\hbar}$, which has been invoked to describe the strange-metal resistivity in overdoped cuprates. In other words, the scaling that governs transport also emerges here in the spectral function, captured entirely by the dynamic scatterer framework.

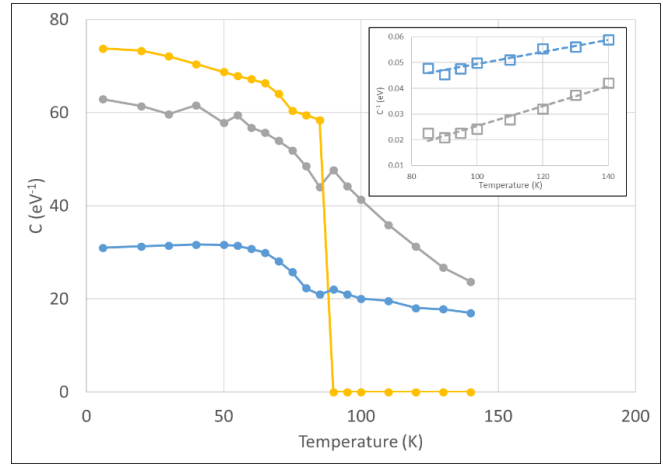

**Figure 11** shows the temperature-dependent values of fitting parameter $C_1$ (gray line and filled gray circles), $C_2$ (yellow line and filled yellow circles), $C_3$ (blue line and filled blue circles). The inset shows the temperature evolution of $C_1^{-1}$ (opened gray squares) and $C_3^{-1}$(opened blue squares). The trend lines (dashed gray and blue lines) show that they are linearly proportional to temperature.

This result strengthens the universality claim. The fact that it appears directly in the linewidth parameter emphasizes that strong scattering, i.e. rather than coherent quasiparticles, is the defining element of the pseudogap phase. Also, transport with hopping channel is shutdown, so that transport resorts to diffusive channel. Without assuming any specific bosonic glue or ordered state, the model reproduces the linear-in-temperature dissipation that defines the strange metal.

### 3.6 Limitations and future outlooks

The present analysis is confined to the antinode spectra of a single overdoped composition. Extension to the node region, where coherent quasiparticles survive above $T_c$, is needed to test whether the same universal functions apply across momentum space. Once, $\rho(E,T)$ can be identified, the corresponding energy, entropy and then free energy can be evaluated, as shown in Figure 12, comparing with those of $\rho_0$. However, the free energy difference between the base system with $\rho_0$ and the true system with $\rho(E,T)$ is very small. This is because this framework is phenomenological in nature and does not identify or include the microscopic origin of superconducting pairing, though it captures the spectral signature of the coherence peak. Nevertheless, the system with $\rho(E,T)$ is already energetically favorable in the superconducting phase. If the pairing mechanism can be identified, the added binding energy will make the system even more favorable. The possible mechanism would be a trapped hole which acts as a binding center of electron pair, or the redistribution density of states provides extra screening to ease pairing. Finally, the ghost peak interpretation demands experimental confirmation, i.e. resolution-dependent ARPES and spectral deconvolution could decisively test whether the peak is interference or a true pole.

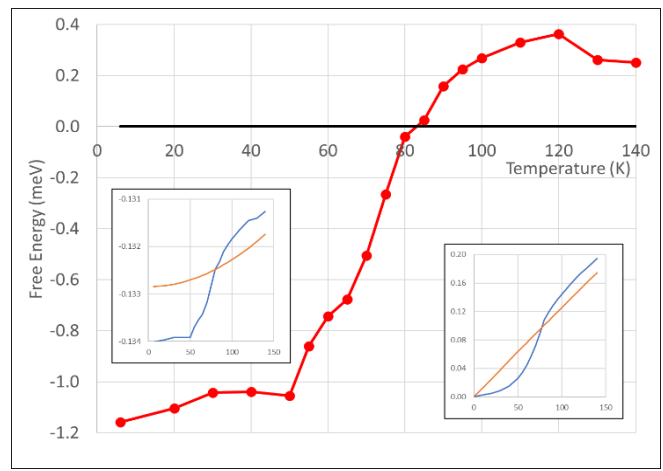

**Figure 12** shows the relative Free energy derived from $\rho(E,T)$ (red line) and $\rho_0$ (black line) as reference. It shows clearly the phase transition. The left inset shows the energy (in eV) from $\rho(E,T)$ (blue line) and $\rho_0$ (orange line). The right inset shows the entropy (in $k_B$) from $\rho(E,T)$ (blue line) and $\rho_0$ (orange line).

Future work should extend the present framework to other doping levels and to different families of superconductors, including nickelates, iron-based systems or recently discovered Kagome-type compounds [21,22], in order to test the universality of the $D_v$ description. The spectral coherence scale $C$ provides a natural bridge to Planckian transport, motivating systematic linewidth studies across materials. If validated, the parabolic-cylinder approach could offer a unifying description for strange metals and high-$T_c$ superconductors, where pseudogap phenomena, coherence peaks, and strange-metal dissipation emerge as distinct realizations of the same universal random system. Connections to the holographic principle—via black-hole analogies—and its reduction to low-dimensional disordered models such as Sachdev–Ye–Kitaev (SYK) model have already provided valuable insights into unconventional superconductivity and strange-metal physics [23,24]. In this picture, the composite parabolic-cylinder function may supply universal solution forms, toward which different theoretical formulations ultimately flow, regardless of their starting assumptions.

**Conclusion**

In this work, I have interpreted the ARPES spectra of over-doped Bi2212 at the antinode within the framework of a particle propagating in a system of dynamic scatterers. The density of states emerges in the analytic form, $\rho(E) = Ae^{-C^2z^2}D_\nu(2Cz)$, with different values of $\nu$ representing different regimes. At high temperature, the spectra resemble the $D_{-\frac{1}{2}}$ solution, typical of a one-dimensional random system. As temperature is reduced, the contributions of $D_{\frac{1}{2}}$ and $D_{\frac{3}{2}}$ gradually evolve, leading to redistribution of states near the Fermi level and the development of the peak–dip feature, which is nothing but the pseudogap. Below $T_c$, the $D_1$ solution emerges, and its amplitude evolves like a superconducting order parameter, saturating at low temperature.

  The inverse of the spectral coherence scale $C$ plays the role of a linewidth parameter. It scales linearly with temperature above $T_c$, consistent with Planckian dissipation in the strange-metal regime. The energy shift $\varepsilon^G$ extracted below $T_c$ slightly underestimates the superconducting gap but tracks the experimental values in a parallel fashion. These results indicate that the essential features of the ARPES spectra, i.e. the hump, the pseudogap, and the coherence peak, can be described within one unified picture of an extremely random system.

  In this picture, the complexity of cuprates is not dismissed but recast, i.e. what appears as exotic many-body physics can also be seen as the universal behavior of electrons in a highly disordered and fluctuating environment. This minimal description does not compete with other theories. Rather, it casts a common ground. Any microscopic model, be it based on strongly correlated interactions, quantum criticality [25] or holographic dualities [23,24], must eventually flow toward this universal regime. The surprising success of parabolic cylinder functions in capturing the spectral evolution across temperature suggests that the strange metal, the pseudogap, and superconductivity itself may be understood as different facets of the same universal description of disorder.

  The present analysis focuses on a single k-point and one doping level of Bi2212 (p = 0.21), yet the framework of parabolic cylinder functions is more general. Future work should extend the fitting to other doping regimes and to the node region, where quasiparticles remain coherent, to test whether the same universal forms can capture both coherent and incoherent spectra. Beyond cuprates, it will be essential to examine other unconventional superconductors, such as iron-based superconductors, nickelates, hydride systems and lower dimension Kagome compounds [21,22], whether they display similar spectral fingerprints of extreme randomness. The ghost peak interpretation also demands further experimental tests, particularly with variable energy resolution, to verify whether the apparent coherence peak is an interference effect or a true quasiparticle pole. Finally, the spectral coherence scale $C$ offers a direct link to Planckian scattering; systematic measurements of linewidths across families and doping levels could reveal whether this scaling is truly universal. In this way, the present work may serve as a starting point for unifying, once thought to be disparate, observations of strange metals and the unconventional high-$T_c$ superconductors under the same minimal description of a particle in a system of dynamic scatterers.

**AI declaration**

While the idea, calculations, and manuscript were entirely composed by me, I acknowledge the use of ChatGPT (OpenAI) for support in editing and refining the presentation of this work. In addition, ChatGPT acted to provide constructive criticism, which, I hope, may help to reduce bias.

**Acknowledgement**

I am deeply grateful to Prof. Worawat Meevasana, whose guidance and insight first brought this problem to my attention. I owe special thanks to my wife, Ekanong Sookasem, for her unwavering support and for continuously providing monthly research funding. I thank Dr. Jakkapat Seeyangnok for introducing the holographic principle and Sachdev-Ye-Kitaev (SYK) model. Finally, I dedicate this work with respect and gratitude to my teachers—Prof. Virulh Sa-yakanit, Prof. Suthat Yoksan, Prof. Wichit Sritrakool, and Prof. Pongkaew Udomsamuthirun—whose wisdom and encouragement have shaped my academic journey.

**Appendix**

**Brief derivation of Green's function**

By using Jensen's inequality, Green's function in Eq. (5) can be obtained by

$$G(t) \approx G_0 \cdot \langle e^{\frac{i(S-S_0)}{\hbar}} \rangle_{S_0} \geq G_0 \exp\left[\frac{i}{\hbar}\langle S - S_0 \rangle_{S_0}\right]. \tag{A1}$$

The trial action in Eq. (6) gives

$$G_0 \exp\left[-\frac{i}{\hbar}\langle S_0 \rangle_{S_0}\right] = \left(\frac{m}{2\pi i \hbar t}\right)^{d/2} \left(\frac{\omega t}{2\sin\frac{\omega t}{2}}\right)^d \exp\left[\frac{d}{2}\left(\frac{\omega t}{2}\cot\frac{\omega t}{2} - 1\right)\right]. \tag{A2}$$

As $t \to 0$, it can be shown that

$$\lim_{t \to 0} G_0 \exp\left[-\frac{i}{\hbar}\langle S_0 \rangle_{S_0}\right] = \left(\frac{m}{2\pi i \hbar t}\right)^{d/2}, \tag{A3}$$

which resembles the normalization factor of a free particle. Also, in statistical physics, the limit $t \to \infty$ is equivalent to the limit $T \to 0$, where only the ground state survives. Thus, the only terms with energy $\frac{\hbar\omega}{2}$ are taken into the account. Consequently,

$$\sin\frac{\omega t}{2} \approx -\frac{i}{2}e^{\frac{i\omega t}{2}}, \tag{A4}$$

$$\cos\frac{\omega t}{2} \approx \frac{1}{2}e^{\frac{i\omega t}{2}}, \tag{A5}$$

and

$$\cot\frac{\omega t}{2} \approx i. \tag{A6}$$

By using Eq. (A4) – (A6), the factor associating with the trial action becomes

$$\lim_{t\to\infty} G_0 \exp\left[-\frac{i}{\hbar}\langle S_0\rangle_{S_0}\right] \approx \left(\frac{m\omega^2}{2\pi\hbar e}\right)^{\frac{d}{2}} (it)^{\frac{d}{2}} \exp\left[-\frac{d}{4}i\omega t\right]. \tag{A7}$$

The most difficult part is the factor associating with the system's action. It can be derived from

$$\frac{i}{\hbar}\langle S\rangle_{S_0} = -\frac{\alpha}{2\hbar^2}\iint d\tau d\sigma \int \frac{d\boldsymbol{q}}{(2\pi)^d} W(q) e^{-B^2 q^2/2}, \tag{A8}$$

where

$$B^2 = \frac{i\hbar}{m}\frac{\sin\frac{\omega\sigma}{2}\sin\frac{\omega(\tau-\sigma)}{2}}{\omega\sin\frac{\omega\tau}{2}}, \tag{A9}$$

with these limiting behaviors,

$$\lim_{t\to 0} B^2 = \frac{i\hbar}{2m}\frac{\sigma(\tau-\sigma)}{\tau}, \tag{A10}$$

and

$$\lim_{t\to\infty} B^2 \approx \frac{\hbar}{2m\omega}. \tag{A11}$$

For Gaussian scatterers, the effective potential can be written as

$$W\big(\boldsymbol{X}(\tau) - \boldsymbol{X}(\sigma)\big) = \frac{i\alpha}{2\hbar}\int \frac{d\boldsymbol{q}}{(2\pi)^d} \exp\left[-\frac{L^2 q^2}{4}\right], \tag{A12}$$

where

$$W(q) = \exp\left[-\frac{L^2 q^2}{4}\right]. \tag{A13}$$

By inserting (A13) into (A8), the factor associating with the system's action can be derived as

$$\frac{i}{\hbar}\langle S\rangle_{S_0} = -\frac{\alpha}{2\hbar^2}\iint d\tau d\sigma \int \frac{d\mathbf{q}}{(2\pi)^d}\exp\left[-\left(\frac{L^2+2B^2}{4}\right)q^2\right], \tag{A14}$$

with these limiting behaviors

$$\lim_{t\to 0}\frac{i}{\hbar}\langle S\rangle_{S_0} \approx -\frac{\alpha t^2}{2\hbar^2(\pi L^2)^{\frac{d}{2}}}, \tag{A15}$$

and

$$\lim_{t\to\infty}\frac{i}{\hbar}\langle S\rangle_{S_0} \approx -\frac{\alpha t^2}{2\hbar^2\left(\pi L^2+\frac{4\pi\hbar}{m\omega}\right)^{\frac{d}{2}}}. \tag{A16}$$

By inserting Eq. (A3) and (A14) into (A1), the solution at the limit $t \to 0$ can be obtained as

$$G(t) = A' \cdot (it)^{-d/2}\exp\left[-\frac{i}{\hbar}\varepsilon^0 t - \frac{t^2}{8C^2}\right], \tag{A17}$$

where

$$A' = \left(\frac{m}{2\pi\hbar}\right)^{d/2}, \tag{A18}$$

and

$$\frac{1}{8C^2} = \frac{\alpha}{2\hbar^2(\pi L^2)^{d/2}}. \tag{A19}$$

It is worth noting that $\varepsilon^0$ has been put by hand, in order to ease band alignment in the experimental data. The parameter $A'$ is closely related, but not the same as $A$.

By inserting Eq. (A7) and (A16) into (A1), the solution at the limit $t \to \infty$ can be obtained as

$$G(t) = A' \cdot (it)^{d/2}\exp\left[-\frac{i}{\hbar}\varepsilon^G t - \frac{t^2}{8C^2}\right], \tag{A20}$$

where

$$A' = \left(\frac{m\omega^2}{2\pi\hbar e}\right)^{d/2}, \tag{A21}$$

and

$$\varepsilon^G = \frac{d}{4}\hbar\omega, \tag{A22}$$

and

$$\frac{1}{8C^2} = \frac{\alpha}{2\hbar^2\left(\pi L^2+\frac{4\pi\hbar}{m\omega}\right)^{\frac{d}{2}}}. \tag{A23}$$

It is worth noting that $A'$ in (A18) and (A21) are closely related to, but not the same as $A$ in Eq. (11) and (13).